\documentstyle[preprint,aps,epsf]{revtex}

\newtheorem{theorem}{Theorem}
\newtheorem{acknowledgement}[theorem]{Acknowledgement}

\begin{document}
\title{Boson Stars with Self-Interacting Quantum Scalar Fields}
\author{Jeongwon Ho$^{1,2}\thanks{%
jwho@phys.ualberta.ca}$, F.C. Khanna$^{1,3}\thanks{%
khanna@phys.ualberta.ca}$, and Chul H. Lee$^{1,4}\thanks{%
chlee@hepth.hanyang.ac.kr}$\\
\vspace{5mm}
\small \vspace{- 3mm} $^{1}$Theoretical Physics Institute, Department of Physics,\\
\small \vspace{- 3mm} University of Alberta, Edmonton, Alberta, Canada T6G 2J1\\
\small \vspace{- 3mm} $^{2}$Deparment of Physics, Kunsan National University, Kunsan 573-701, Korea%
\\
\small \vspace{- 3mm} $^{3}$TRIUMF, 4004 Wesbrook Mall, Vancouver, British Columbia, Canada V6T 2A3%
\\
\small \vspace{- 3mm} $^{4}$Department of Physics, Hanyang University, Seoul 133-791, Korea}
\date{\today}
\maketitle

\begin{abstract}
The Klein-Gordon-Einstein equations of classical real scalar fields have
time-dependent solutions (periodic in time). We show that quantum real
scalar fields can form {\it non-oscillating (static) }solitonic objects,
which are quite similar to the solutions describing boson stars formed with
classical and quantum complex scalar fields (the latter will be studied in
this paper). We numerically analyze the difference between them concerning
the mass of boson stars. On the other hand, we suggest an interesting test
(a viable process that the boson star may undergo in the early universe) for
the formation of boson stars. That is, it is questioned that after a
second-order phase transition (a simple toy model will be considered here),
what is the fate of the boson star composed of quantum real scalar field.
\end{abstract}

\section{Introduction}

The presence of dark matter has been established indirectly in a wide range
of scale of the universe, from that of individual galaxies to the entire
universe itself \cite{silk}. Though direct measurements of the nature of the
dark matter have not yielded any result, speculations on its composition
vary from baryonic to non-baryonic matter. Particles like axions and
neutralinos are the specific targets for direct observation since the
indirect measurements like rotation curves of spiral galaxies and others do
not depend on the presence of a particular type of particles. One of the
most promising candidates for dark matter is the boson star, which was
discovered theoretically over thirty years ago \cite{kaup}\cite{ruffini}.
Until now, the reality of boson stars has been successfully applied to
various plausible physical situations \cite{mielke}.

The boson star is a self-gravitating compact solitonic object made up of
bosonic fields. Non-interacting complex scalar fields \cite{kaup}\cite%
{ruffini} were originally considered for the constituents composing boson
stars. In this case, the resultant configurations are typically `mini'-boson
stars, which have small size and mass. This result originates from the
following specific feature of boson stars; the boson star is protected from
gravitational collapse by the Heisenberg uncertainty principle, instead of
the Pauli exclusion principle that applies to fermionic stars, and the
characteristic length scale of the former is much smaller than that of the
latter. It has been shown that this situation can be dramatically changed by
introducing self-interacting complex scalar fields. The self-interaction
effectively generates a repulsive force and the maximum mass of stable boson
stars can be enhanced up to a size of the order of ordinary fermionic stars %
\cite{colpi}\footnote{%
Analyltic evaluation of the maximum mass and higher order self-interaction
effect to boson star configuration can be found in \cite{jwho}.}.

On the other hand, there is another type of gravitationally bound solitonic
object known as oscillating soliton star composed of classical real scalar
fields \cite{seidel}. In this case, the spacetime geometry and real scalar
field satisfying the Klein-Gordon-Einstein (KGE) equations are
time-dependent (periodic in time). Such objects are of interest to study
dark matter, since the most promising candidates for dark matter are
described by real scalar fields such as the axion \cite{kim}. However, even
though these solutions are stable with respect to some simple perturbations,
it is still unclear whether the stability of the oscillating star can be
maintained with general perturbations. Moreover, due to its oscillating
feature, real scalar fields may not form primordial solitonic objects in the
early universe \cite{seidel2} (cf. \cite{kolb}), which are expected to play
important roles in galaxy formation, the microwave background, and formation
of protostars.

In this paper, we report some very interesting results for the
self-gravitating solitonic object formed with real scalar fields: there
exist gravitationally bound {\it non-oscillating (static) }objects composed
of {\it quantum real }scalar fields, instead of {\it classical real} ones.
Our (zero-node) solitonic solution is quite similar to the solutions
describing boson stars formed with classical \cite{colpi} and quantum
complex scalar fields (the latter will be studied in this paper). The only
difference between them at the equation of motion level is just the
effective coefficient of the self-interaction term (within the validity
considered in this paper, i.e. zero-node solutions, the semi-classical and
Hartree (mean-field) approximations). Therefore, non-self-interacting
quantum/classical complex and quantum real scalar fields coupled to gravity
can form identical static mini-boson stars.

On the other hand, as many viable cosmological scenarios indicate, quantum
fields present in early stage of the universe may experience phase
transitions by temperature changes. Thus, if the boson star forms in the
early universe, it would undergo phase transitions. It might be an
interesting issue to consider the consequence of such a phase transition for
the boson star.\footnote{In \cite{torres}, a boson star in an evolving
cosmological background with time varying gravitational constant (i.e.,
in the context of scalar-tensor theory of gravity) has been considered.}
What is the fate of the scalar field and what are the final
products after the phase transition, which is presumably second order? In
this paper, as a first step, we consider a simple toy model; without
speculating on the detailed procedure of the phase transition, we introduce
a plausible form of the Lagrangian for real scalar fields coupled to
gravity, which would give the physics after phase transition, and examine
that there may still exist boson stars. If they continue to exist, what are
the differences between the boson stars before and after phase transition.

In the next section, we start by studying boson stars composed of {\it %
quantum} complex scalar fields, and compare them with the boson stars formed
with the {\it classical }complex scalar fields \cite{colpi}. The boson stars
composed of quantum real scalar fields are studied in Sect.3. In Sect.4, we
consider boson stars composed of quantum real scalar fields that undergo a
second-order phase transition. Disscussions and some remarks are given in
Sect.5.

\section{Complex quantum scalar fields}

Consider a self-interacting complex scalar field Lagrangian given by

\begin{equation}
{\cal L}_{{\rm matt}}=-\frac{1}{2}\partial ^{\mu }\phi ^{\ast }\partial
_{\mu }\phi -\frac{1}{2}m^{2}\left| \phi \right| ^{2}-\frac{\lambda }{4}%
\left| \phi \right| ^{4},  \label{lagran}
\end{equation}%
where we assume that the coupling to gauge fields is negligible. Then, the
field equations of the scalar field are given by%
\begin{equation}
\nabla ^{\mu }\nabla _{\mu }\phi -m^{2}\phi -\lambda \left| \phi \right|
^{2}\phi =0,  \label{fdeq}
\end{equation}%
and its complex conjugate. For convenience, we introduce two real scalar
fields, defined by%
\begin{equation}
\Phi_1 =\frac{1}{\sqrt{2}}\left( \phi +\phi ^{\ast }\right) ,\hspace{0.1in}%
\Phi_2 =\frac{i}{\sqrt{2}}\left( \phi -\phi ^{\ast }\right) ,  \label{dffds}
\end{equation}%
and rewrite the field equations as%
\begin{eqnarray}
\nabla ^{\mu }\nabla _{\mu }\Phi_1 -m^{2}\Phi_1 -\lambda \left( \Phi_1^{2}
+\Phi_2^{2}\right) \Phi_1 &=&0,  \label{nfdeq} \\
\nabla ^{\mu }\nabla _{\mu }\Phi_2 -m^{2}\Phi_2 -\lambda \left( \Phi_1^{2}
+\Phi_2^{2}\right) \Phi_2 &=&0.  \nonumber
\end{eqnarray}

General expressions for solutions to Eqs. (\ref{nfdeq}) can be written as%
\begin{eqnarray}
\Phi_1 &=&\sum_{nlm}\left( a_{nlm}\zeta (r)_{nl}Y_{m}^{l}e^{-i\omega
_{nl}^{\zeta }t}+a_{nlm}^{\ast }\zeta (r)_{nl}^{\ast }(Y_{m}^{l})^{\ast
}e^{i\omega _{nl}^{\zeta }t}\right) ,  \label{qtfd} \\
\Phi_2 &=&\sum_{nlm}\left( b_{nlm}\eta (r)_{nl}Y_{m}^{l}e^{-i\omega
_{nl}^{\eta }t}+b_{nlm}^{\ast }\eta (r)_{nl}^{\ast }(Y_{m}^{l})^{\ast
}e^{i\omega _{nl}^{\eta }t}\right) .  \nonumber
\end{eqnarray}%
As quantum scalar fields, $\Phi_1 $ and $\Phi_2 $ become operators%
\begin{eqnarray}
\widehat{\Phi}_1 &=&\sum_{nlm}\left( \widehat{a}_{nlm}\zeta
(r)_{nl}Y_{m}^{l}e^{-i\omega _{nl}^{\zeta }t}+\widehat{a}_{nlm}^{\dagger
}\zeta (r)_{nl}^{\ast }(Y_{m}^{l})^{\ast }e^{i\omega _{nl}^{\zeta }t}\right)
,  \label{qtfa} \\
\widehat{\Phi}_2 &=&\sum_{nlm}\left( \widehat{b}_{nlm}\eta
(r)_{nl}Y_{m}^{l}e^{-i\omega _{nl}^{\eta }t}+\widehat{b}_{nlm}^{\dagger
}\eta (r)_{nl}^{\ast }(Y_{m}^{l})^{\ast }e^{i\omega _{nl}^{\eta }t}\right) ,
\nonumber
\end{eqnarray}%
where $\widehat{a}_{nlm}^{\dagger }$, $\widehat{b}_{nlm}^{\dagger }$ and $%
\widehat{a}_{nlm}$, $\widehat{b}_{nlm}$ are the creation and annihilation
operators, respectively, and they satisfy the commutation relations $[%
\widehat{a}_{\alpha }^{\dagger },\widehat{a}_{\alpha ^{\prime }}]=$ $[%
\widehat{b}_{\alpha }^{\dagger },\widehat{b}_{\alpha ^{\prime }}]=\delta
_{\alpha \alpha ^{\prime }}$. In the following, only the ground state of the
system, $n=1,l=m=0$, will be considered and the condition $\omega
_{01}^{\Phi_1}=\omega _{01}^{\Phi_2}=\omega $ is assumed. In addition, we
assume that in the ground state, which is denoted by $|N>=|N_{\Phi_1
}>\otimes |N_{\Phi_2 }>$, the number of particles created by $\widehat{a}%
^{\dagger }$ is the same as that associated with $\widehat{b}^{\dagger }$,
i.e. $N_{\Phi_1}=N_{\Phi_2}=N$. In general, $\zeta (r)$ and $\eta (r)$ are
complex functions, but here we restrict them to real ones. To linearize the
field equations (\ref{nfdeq}), we adopt the Hartree approximation such that $%
\left( \widehat{\Phi}_1^{2}+\widehat{\Phi}_2^{2}\right) \widehat{\Phi}_1\sim
<N|(\widehat{\Phi}_1^{2}+\widehat{\Phi}_2^{2})|N>\widehat{\Phi}_1$.

On the other hand, we also use the semi-classical approximation in
Einstein's equations%
\begin{equation}
G_{\nu }^{\mu }=8\pi G<N|\widehat{T}_{\nu }^{\mu }|N>,  \label{eineq}
\end{equation}%
where $G_{\nu }^{\mu }$ is the Einstein tensor and $\widehat{T}_{\nu }^{\mu
} $ is the energy-momentum operator that is obtained by quantizing the
energy-momentum tensor%
\begin{eqnarray}
T_{\nu }^{\mu } &=&\frac{1}{2}\left( \partial ^{\mu }\phi ^{\ast }\partial
_{\nu }\phi +\partial ^{\mu }\phi \partial _{\nu }\phi ^{\ast }\right) -%
\frac{1}{2}g_{\nu }^{\mu }\left( \partial ^{\alpha }\phi ^{\ast }\partial
_{\alpha }\phi +m^{2}\left| \phi \right| ^{2}+\frac{\lambda }{2}\left| \phi
\right| ^{4}\right)  \label{emtensor} \\
&=&\frac{1}{2}\left( \partial ^{\mu }\Phi_1 \partial _{\nu }\Phi_1 +\partial
^{\mu }\Phi_2 \partial _{\nu }\Phi_2 \right) -\frac{1}{4}g_{\nu }^{\mu }\left(
\partial ^{\alpha }\Phi_1 \partial _{\alpha }\Phi_1 +\partial ^{\alpha }\Phi_2
\partial _{\alpha }\Phi_2 +m^{2}\left( \Phi_1^{2}+\Phi_2 ^{2}\right) +\frac{%
\lambda }{4}\left( \Phi_1^{2}+\Phi_2^{2}\right)^{2}\right) .  \nonumber
\end{eqnarray}%
The background spacetime is restricted to be static and spherically
symmetric. Thus, in Schwarzschild coordinates, it can be written as%
\begin{equation}
ds^{2}=-B(r)dt^{2}+A(r)dr^{2}+r^{2}d\Omega _{2}^{2},  \label{metric}
\end{equation}%
where $d\Omega _{2}^{2}$ denotes the metric of a two-dimensional unit sphere.

After performing some algebraic calculations and reparametrizations in Eqs. (%
\ref{eineq}) and (\ref{nfdeq}), we obtain the equations%
\begin{eqnarray}
&&{\frac{A^{\prime }}{A^{2}x}}+{\frac{1}{x^{2}}}\left( 1-{\frac{1}{A}}%
\right) =\left( {\frac{\Omega ^{2}}{B}}+1\right) (\sigma _{1}^{2}+\sigma
_{2}^{2})+{\frac{\Lambda }{4}}\left( 3\left( \sigma _{1}^{4}+\sigma
_{2}^{4}\right) +4\sigma _{1}^{2}\sigma _{2}^{2}\right) +{\frac{1}{A}}\left(
\left( \sigma _{1}^{\prime }\right) ^{2}+\left( \sigma _{2}^{\prime }\right)
^{2}\right) ,  \nonumber \\
&&{\frac{B^{\prime }}{ABx}}-{\frac{1}{x^{2}}}\left( 1-{\frac{1}{A}}\right)
=\left( {\frac{\Omega ^{2}}{B}}-1\right) (\sigma _{1}^{2}+\sigma _{2}^{2})-{%
\frac{\Lambda }{4}}\left( 3\left( \sigma _{1}^{4}+\sigma _{2}^{4}\right)
+4\sigma _{1}^{2}\sigma _{2}^{2}\right) +{\frac{1}{A}}\left( \left( \sigma
_{1}^{\prime }\right) ^{2}+\left( \sigma _{2}^{\prime }\right) ^{2}\right) ,
\nonumber \\
&&\sigma _{1}^{\prime \prime }+\left( {\frac{2}{x}}+{\frac{B^{\prime }}{2B}}-%
{\frac{A^{\prime }}{2A}}\right) \sigma _{1}^{\prime }+A\left[ \left( {\frac{%
\Omega ^{2}}{B}}-1\right) \sigma _{1}-\Lambda (\sigma _{1}^{2}+\sigma
_{2}^{2})\sigma _{1}\right] =0,  \nonumber \\
&&\sigma _{2}^{\prime \prime }+\left( {\frac{2}{x}}+{\frac{B^{\prime }}{2B}}-%
{\frac{A^{\prime }}{2A}}\right) \sigma _{2}^{\prime }+A\left[ \left( {\frac{%
\Omega ^{2}}{B}}-1\right) \sigma _{2}-\Lambda (\sigma _{1}^{2}+\sigma
_{2}^{2})\sigma _{2}\right] =0,  \label{eqmx}
\end{eqnarray}%
where $x\equiv mr$, $\sigma _{1}\equiv \left( 4\pi GN\right) ^{1/2}\zeta (r)$%
, $\sigma _{2}\equiv \left( 4\pi GN\right) ^{1/2}\eta (r)$, $\Omega \equiv {%
\omega /m}$, $\Lambda \equiv \lambda /4\pi Gm^{2}$, and the prime denotes
the derivative with respect to $x$. In Eqs. (\ref{eqmx})\footnote{Note that
while Eqs.(\ref{nfdeq}) and (\ref{emtensor})
are invariant under a rotation in the internal space given by
$\Phi_1 \rightarrow \Phi_1 \cos\Theta + \Phi_2 \sin\Theta$, $\Phi_2
\rightarrow -\Phi_1 \sin\Theta + \Phi_2 \cos\Theta$, where $\Theta $ is
a constant, in Eq.(\ref{eqmx})
such a symmetry does not exist. This is simply because $\sigma_1$, $\sigma_2$
in Eq.(\ref{eqmx}) are not field variables but (redefined) radial functions
of the mode solutions. So, there is no reason maintaining the internal
symmetry in the level of Eq.(\ref{eqmx}).}, we have used the
approximation, $N\gg 1$. In this paper, we consider only a special type of
solution given by $\sigma _{1}=\sigma _{2}=\sigma /\sqrt{2}$. Then, the
equations become%
\begin{eqnarray}
&&{\frac{A^{\prime }}{A^{2}x}}+{\frac{1}{x^{2}}}\left( 1-{\frac{1}{A}}%
\right) =\left( {\frac{\Omega ^{2}}{B}}+1\right) \sigma ^{2}+{\frac{5\Lambda 
}{8}}\sigma ^{4}+{\frac{\left( \sigma ^{\prime }\right) ^{2}}{A}},
\label{eqm1} \\
&&{\frac{B^{\prime }}{ABx}}-{\frac{1}{x^{2}}}\left( 1-{\frac{1}{A}}\right)
=\left( {\frac{\Omega ^{2}}{B}}-1\right) \sigma ^{2}-{\frac{5\Lambda }{8}}%
\sigma ^{4}+{\frac{\left( \sigma ^{\prime }\right) ^{2}}{A}}.  \label{eqm2}
\\
&&\sigma ^{\prime \prime }+\left( {\frac{2}{x}}+{\frac{B^{\prime }}{2B}}-{%
\frac{A^{\prime }}{2A}}\right) \sigma ^{\prime }+A\left[ \left( {\frac{%
\Omega ^{2}}{B}}-1\right) \sigma -\Lambda \sigma ^{3}\right] =0  \label{eqm3}
\end{eqnarray}%
Note that comparing (\ref{eqm1})-(\ref{eqm3}) with the KGE equations for 
{\it classical }complex scalar fields \cite{colpi}, only the coefficients of
self-interaction terms in (\ref{eqm1}) and (\ref{eqm2}) are changed as $%
\Lambda /2\rightarrow 5\Lambda /8$, while the field equation (\ref{eqm3})
remains without any change. Such a change results in modifications of some
properties of the boson stars, especially the maximum mass of stable boson
stars. This can be seen in the following numerical analysis.

Substituting an ansatz of the metric function $A$ by%
\begin{equation}
A\left( x\right) =\left[ 1-{\frac{2{\cal M}\left( x\right) }{x}}\right]
^{-1},  \label{aform}
\end{equation}%
into the above equations and using the boundary conditions%
\begin{equation}
{\cal M}(0)=0,\hspace{0.06in}\sigma (0)=\sigma _{c},\hspace{0.06in}\sigma
^{^{\prime }}(0)=0,\hspace{0.06in}{\rm and\hspace{0.06in}}B(\infty )=1,
\label{bdcd}
\end{equation}%
we obtain the mass of boson stars as a function of $\sigma _{c}$ for $%
\Lambda =100$ in Fig.1. In Fig.1, it is shown that the maximum mass of boson
stars composed of quantum complex scalar fields is greater than that in the
case of classical complex scalar fields. This is true for other values of $%
\Lambda $ as can be seen in Fig.2. The lines in Fig.2 are chosen in a way
that they hold for $\Lambda \gg 1$; in the case of quantum fields
($5\Lambda /8$) $M_{\max}\approx 0.223\Lambda ^{1/2}M_{p}^{2}/m$, while
$M_{\max }\approx 0.22\Lambda ^{1/2}M_{p}^{2}/m$ for the classical fields
($\Lambda /2$) \cite{colpi}. It has
also to be mentioned that in Fig.1, there is a crossing (at $\sigma
_{c}\approx 0.132$) of the two mass-curves, and when $\sigma _{c}>0.132$,
the mass of the boson stars formed by the classical complex scalar field is
greater than that of the boson stars formed by a quantum complex scalar
field.

As mentioned above, the difference between the two cases of classical and
quantum fields is in the coefficient of the self-interaction term in the
Einstein equations (\ref{eqm1}) and (\ref{eqm2}), i.e. $\Lambda /2$ for the
classical field and $5\Lambda /8$ for the quantum one. Why does the
difference appear? In fact, the KGE equations for the classical complex
scalar field considered in Ref.\cite{colpi} become the equations for a
``real'' field\footnote{%
The quoted term of ``real'' field will be used for the spatial part of a
mode function, not for an ordinary field.} after eliminating the
time-dependent part. This is because the authors of Ref.\cite{colpi} have
taken the solution of the complex field which has the form of
$\phi (r,t)=\Phi (r)\exp (-i\omega t)$ with real $\Phi (r)$. In our
case, the KGE equations given in (\ref{eqmx}) are for two ``real'' fields.
Even though we chose a special form of the solution given by $\sigma
_{1}=\sigma _{2}=\sigma /\sqrt{2}$ to compare with the classical case
studied in \cite{colpi}, the resulting equations given by (\ref{eqm1})-(\ref%
{eqm3}) already include the interaction between the two ``real'' fields as
well as self-interactions of each fields, and this is the reason why the
coefficients of self-interaction term are {\it effectively }different from
each other in the quantum and classical cases, even though we have started
with the same Largrangian (\ref{lagran}).

Roughly speaking, since larger self-interaction energy exerts greater
repulsive force in the formation of boson stars, the `quantum' effect, i.e.
using the quantum field rather than the classical one to form a boson star,
enhances the maximum mass of the boson stars. (In fact, when the
self-interaction is introduced, density of the boson star is approximately
proportional to $\Lambda ^{-1}$ so that the boson stars become dilute.
However, since its radius is approximately proportional to $\Lambda ^{1/2}$,
the maximum mass of the boson star is to be proportional to $\Lambda ^{1/2}$%
. This argument is available for the case that the self-interaction energy
is comparable to the mass energy of the field, i.e. $\Lambda >>0$. See Ref.%
\cite{jwho} for a detailed analysis.)

Of course, the validity of our argument has to be restricted to solutions
that are considered in this paper and \cite{colpi}. However, it is
reasonable to study the difference between the roles of classical and
quantum fields in the formation of boson stars. Another point that has to be
mentioned is that the mass curves in Fig.1 cross each other at $\sigma
_{c}\approx 0.132$, and the boson star masses for the classical and quantum
cases reverse in magnitude. This phenomenon can not be described by the
above argument. To understand such a behavior, a detailed analytical
approach would be required.

\section{Real quantum scalar fields}

Now, let us consider a real scalar field for the formation of boson stars.
The real massive scalar field Lagrangian including self-interaction is
written as%
\begin{equation}
{\cal L}_{{\rm matt}}=-\frac{1}{2}\partial ^{\mu }\Phi \partial _{\mu }\Phi -%
\frac{1}{2}m^{2}\Phi ^{2}-\frac{\lambda }{4}\Phi ^{4}.  \label{lagbf}
\end{equation}%
And the energy-momentum tensor and field equation are obtained as%
\begin{eqnarray}
&T_{\nu }^{\mu }=\partial ^{\mu }\Phi \partial _{\nu }\Phi -\frac{1}{2}%
g_{\nu }^{\mu }\left( \partial ^{\alpha }\Phi \partial _{\alpha }\Phi
+m^{2}\Phi ^{2}+\frac{\lambda }{2}\Phi ^{4}\right) ,&  \label{emtrl} \\
&\nabla ^{\mu }\nabla _{\mu }\Phi -m^{2}\Phi -\lambda \Phi ^{3}=0.&
\label{eqrl}
\end{eqnarray}

From the general form of solutions of the field equation, the field operator
is given by 
\begin{equation}
\widehat{\Phi }=\sum_{nlm}\left( \widehat{d}_{nlm}\Phi
(r)_{nl}Y_{m}^{l}e^{-i\omega _{nl}t}+\widehat{d}_{nlm}^{\dagger }\Phi ^{\ast
}(r)_{nl}(Y_{m}^{l})^{\ast }e^{i\omega _{nl}t}\right) ,  \label{realsc}
\end{equation}%
where the creation and annihilation operators, $\widehat{d}_{nlm}^{\dagger }$
and $\widehat{d}_{nlm}$, satisfy the commutation relations $[\widehat{d}%
_{\alpha }^{\dagger },\widehat{d}_{\alpha ^{\prime }}]=\delta _{\alpha
\alpha ^{\prime }}$. Again we only consider the ground state. Note that we
take the function $\Phi (r)$ as a complex function. Then, defining new real
functions%
\begin{equation}
\xi (r)\equiv \frac{1}{\sqrt{2}}\left( \Phi (r)+\Phi ^{\ast }(r)\right) ,%
\hspace{0.1in}\chi (r)\equiv \frac{i}{\sqrt{2}}\left( \Phi (r)-\Phi ^{\ast
}(r)\right) ,  \label{dfmdftn}
\end{equation}%
and taking the Hartree and semi-classical approximations, we obtain the KGE
equations given by%
\begin{eqnarray}
&{\frac{A^{\prime }}{A^{2}x}}+{\frac{1}{x^{2}}}\left( 1-{\frac{1}{A}}\right)
=\left( {\frac{\Omega ^{2}}{B}}+1\right) (\rho _{1}^{2}+\rho _{2}^{2})+{%
\frac{3\Lambda }{4}}\left( \rho _{1}^{2}+\rho _{2}^{2}\right) ^{2}+{\frac{1}{%
A}}\left( \left( \rho _{1}^{\prime }\right) ^{2}+\left( \rho _{2}^{\prime
}\right) ^{2}\right) ,&  \nonumber \\
&{\frac{B^{\prime }}{ABx}}-{\frac{1}{x^{2}}}\left( 1-{\frac{1}{A}}\right)
=\left( {\frac{\Omega ^{2}}{B}}-1\right) (\rho _{1}^{2}+\rho _{2}^{2})-{%
\frac{3\Lambda }{4}}\left( \rho _{1}^{2}+\rho _{2}^{2}\right) ^{2}+{\frac{1}{%
A}}\left( \left( \rho _{1}^{\prime }\right) ^{2}+\left( \rho _{2}^{\prime
}\right) ^{2}\right) ,&  \nonumber \\
&\rho _{1}^{\prime \prime }+\left( {\frac{2}{x}}+{\frac{B^{\prime }}{2B}}-{%
\frac{A^{\prime }}{2A}}\right) \rho _{1}^{\prime }+A\left[ \left( {\frac{%
\Omega ^{2}}{B}}-1\right) \rho _{1}-\Lambda (\rho _{1}^{2}+\rho
_{2}^{2})\rho _{1}\right] =0,&  \nonumber \\
&\rho _{2}^{\prime \prime }+\left( {\frac{2}{x}}+{\frac{B^{\prime }}{2B}}-{%
\frac{A^{\prime }}{2A}}\right) \rho _{2}^{\prime }+A\left[ \left( {\frac{%
\Omega ^{2}}{B}}-1\right) \rho _{2}-\Lambda (\rho _{1}^{2}+\rho
_{2}^{2})\rho _{2}\right] =0,&  \label{eeq}
\end{eqnarray}%
where $x\equiv mr$, $\rho _{1}\equiv \left( 4\pi GN\right) ^{1/2}\xi (r)$, $%
\rho _{2}\equiv \left( 4\pi GN\right) ^{1/2}\chi (r)$, $\Omega \equiv {%
\omega /m}$, $\Lambda \equiv \lambda /4\pi Gm^{2}$, and we use the metric (%
\ref{metric}) and the approximation $N\gg 1$. Taking a special form of the
solution for $\rho _{1}=\rho _{2}=\sigma /\sqrt{2}$, above equations become%
\begin{eqnarray}
&{\frac{A^{\prime }}{A^{2}x}}+{\frac{1}{x^{2}}}\left( 1-{\frac{1}{A}}\right)
=\left( {\frac{\Omega ^{2}}{B}}+1\right) \sigma ^{2}+{\frac{3\Lambda }{4}}%
\sigma ^{4}+{\frac{\left( \sigma ^{\prime }\right) ^{2}}{A}},&  \label{eeq1}
\\
&{\frac{B^{\prime }}{ABx}}-{\frac{1}{x^{2}}}\left( 1-{\frac{1}{A}}\right)
=\left( {\frac{\Omega ^{2}}{B}}-1\right) \sigma ^{2}-{\frac{3\Lambda }{4}}%
\sigma ^{4}+{\frac{\left( \sigma ^{\prime }\right) ^{2}}{A}},&  \label{eq2}
\\
&\sigma ^{\prime \prime }+\left( {\frac{2}{x}}+{\frac{B^{\prime }}{2B}}-{%
\frac{A^{\prime }}{2A}}\right) \sigma ^{\prime }+A\left[ \left( {\frac{%
\Omega ^{2}}{B}}-1\right) \sigma -\Lambda \sigma ^{3}\right] =0.&
\label{feq}
\end{eqnarray}%
Note that the above equations can be obtained from (\ref{eqmx}) by
substituting $\sigma _{1}=\sigma $, $\sigma _{2}=0$.

Comparing Eqs. (\ref{eeq1})-(\ref{feq}) with (\ref{eqm1})-(\ref{eqm3}), we
find that the only difference is the coefficient of the self-interaction
term. Thus, the KGE equations (\ref{eeq1})-(\ref{feq}) should have solutions
that are much like the solutions describing the boson star composed of
classical/quantum complex scalar fields. As mentioned above, since the
coefficient $3\Lambda /4$ is greater than those in the classical and quantum
complex scalar field cases, which are $\Lambda /2$ and $5\Lambda /8$,
respectively, the maximum mass of boson star formed by quantum real scalar
fields should be larger than those of other cases. Indeed, using the ansatz (%
\ref{aform}) and the boundary conditions given by (\ref{bdcd}), we calculate
numerically the mass of boson star as a function of $\sigma _{c}$ for
various values of $\Lambda $ in Fig.3. In Fig.1 and Fig.2, we compare the
maximum mass of boson stars composed of quantum real scalar fields with
other cases, and confirm that the order of magnitude of maximum masses
follows that of the coefficient of self-interaction term, i.e. $3\Lambda
/4>5\Lambda /8>\Lambda /2$. (The solid line in Fig.2 is the relation
$M_{\max}\approx 0.225\Lambda ^{1/2}M_{p}^{2}/m$, which holds for $\Lambda
>> 1$ for the case of quantum real scalar fields.)
This result can be interpreted in the same way
as given in Sect.2; comparing (\ref{eeq}) with (\ref{eqmx}), the Einstein
equations in (\ref{eeq}) include the term representing the interaction of
``real'' fields, $(3\Lambda /2)\rho _{1}^{2}\rho _{2}^{2}$, which has a
larger coefficient than in other cases, e.g. $\Lambda \sigma _{1}^{2}\sigma
_{2}^{2}$ in the Einstein equations in (\ref{eqmx}).

\section{Phase transition}

In this section, we consider a phase transition that the boson star can
undergo in early stages of the universe. Our model of the system after the
phase transition is a simple one with Lagrangian for real scalar fields
given by%
\begin{equation}
{\cal L}_{{\rm matt}}^{ph}=-\frac{1}{2}\partial ^{\mu }\Phi \partial _{\mu
}\Phi +\frac{1}{2}m^{2}\Phi ^{2}-\frac{\lambda }{4}\Phi ^{4}.  \label{lagaf}
\end{equation}%
In order to simulate the condition of phase transition the mass $m$ in the
Largrangian is taken to be a function of time $t$. Then, in the process of
the phase transition, we consider $m^{2}\rightarrow -m^{2}$ after phase
transition. Such a procedure is commonly used in describing a second-order
phase transition, and compares well to the ideas of spontaneous symmetry
breaking which is the theoretical underpinning for such a phase transition.

In this theory, the vacuum expectation value of the field does not vanish,
but $<\Phi >_{vac}=m\sqrt{\lambda }$. Now, shifting the field, $\Phi
\rightarrow \Phi +m/\sqrt{\lambda }$, and the potential energy, $V(\Phi
)\rightarrow V(\Phi )-m^{4}/(4\lambda )$, we obtain a new Lagrangian written
as%
\begin{equation}
{\cal L}_{{\rm matt}}^{new}=-\frac{1}{2}\partial ^{\mu }\Phi \partial _{\mu
}\Phi -m^{2}\Phi ^{2}-\frac{\lambda }{4}\Phi ^{4}-m\sqrt{\lambda }\Phi ^{3}.
\label{newlag}
\end{equation}%
Then, the energy-momentum tensor and field equation are evaluated from the
Lagrangian (\ref{newlag});%
\begin{eqnarray}
&T_{\nu }^{\mu }=\partial ^{\mu }\Phi \partial _{\nu }\Phi -\frac{1}{2}%
g_{\nu }^{\mu }\left( \partial ^{\alpha }\Phi \partial _{\alpha }\Phi
+2m^{2}\Phi ^{2}+\frac{\lambda }{2}\Phi ^{4}+2m\sqrt{\lambda }\Phi
^{3}\right) ,&  \label{newemt} \\
&\nabla ^{\mu }\nabla _{\mu }\Phi -2m^{2}\Phi -\lambda \Phi ^{3}-3m\sqrt{%
\lambda }\Phi ^{2}=0.&  \label{newfe}
\end{eqnarray}%
Comparing (\ref{newemt}) and (\ref{newfe}) with (\ref{emtrl}) and (\ref{eqrl}%
) for the case before the phase transition, we see that the scalar field
becomes effectively more massive, $m\longrightarrow \sqrt{2}m$, through the
phase transition, and an additional non-linear interaction term with a
coupling constant proportional to $m\sqrt{\lambda }$ is included in the
energy-momentum tensor and field equation. Within the validity of the
Hartree and semiclassical approximation, however, the non-linear coupling
term proportional to $m\sqrt{\lambda }$ does not contribute to the KGE
equations, i.e. $<N|\widehat{\Phi }^{3}|N>=0$ in the Einstein equations and $%
<N|\widehat{\Phi }|N>\widehat{\Phi }=0$ in the field equation due to the
semiclassical and Hartree approximation, respectively. Thus, effectively the
consequence of phase transition is just to enhance the mass of scalar fields.

Explicitly, the KGE equations are written as%
\begin{eqnarray}
&{\frac{A^{\prime }}{A^{2}x}}+{\frac{1}{x^{2}}}\left( 1-{\frac{1}{A}}\right)
=\left( {\frac{\Omega ^{2}}{B}}+2\right) \sigma ^{2}+{\frac{3\Lambda }{4}}%
\sigma ^{4}+{\frac{\left( \sigma ^{\prime }\right) ^{2}}{A}},&  \label{eq1af}
\\
&{\frac{B^{\prime }}{ABx}}-{\frac{1}{x^{2}}}\left( 1-{\frac{1}{A}}\right)
=\left( {\frac{\Omega ^{2}}{B}}-2\right) \sigma ^{2}-{\frac{3\Lambda }{4}}%
\sigma ^{4}+{\frac{\left( \sigma ^{\prime }\right) ^{2}}{A}},&  \label{eq2af}
\\
&\sigma ^{\prime \prime }+\left( {\frac{2}{x}}+{\frac{B^{\prime }}{2B}}-{%
\frac{A^{\prime }}{2A}}\right) \sigma ^{\prime }+A\left[ \left( {\frac{%
\Omega ^{2}}{B}}-2\right) \sigma -\Lambda \sigma ^{3}\right] =0.&
\label{eq3af}
\end{eqnarray}%
More conveniently, reparametrizing the above equations by $y\equiv \sqrt{2}x$%
, $\overline{\Omega }\equiv \Omega /\sqrt{2}$, we obtain%
\begin{eqnarray}
&{\frac{A^{\prime }}{A^{2}y}}+{\frac{1}{y^{2}}}\left( 1-{\frac{1}{A}}\right)
=\left( {\frac{\overline{\Omega }^{2}}{B}}+1\right) \sigma ^{2}+{\frac{%
3\Lambda }{8}}\sigma ^{4}+{\frac{\left( \sigma ^{\prime }\right) ^{2}}{A},}&
\label{afeq1} \\
&{\frac{B^{\prime }}{ABy}}-{\frac{1}{y^{2}}}\left( 1-{\frac{1}{A}}\right)
=\left( {\frac{\overline{\Omega }^{2}}{B}}-1\right) \sigma ^{2}-{\frac{%
3\Lambda }{8}}\sigma ^{4}+{\frac{\left( \sigma ^{\prime }\right) ^{2}}{A}},&
\label{afeq2} \\
&\sigma ^{\prime \prime }+\left( {\frac{2}{y}}+{\frac{B^{\prime }}{2B}}-{%
\frac{A^{\prime }}{2A}}\right) \sigma ^{\prime }+A\left[ \left( {\frac{%
\overline{\Omega }^{2}}{B}}-1\right) \sigma -\frac{\Lambda }{2}\sigma ^{3}%
\right] =0,&  \label{afeq3}
\end{eqnarray}%
where prime denotes the derivative with respect to $y$. Note that the
coefficient of the self-interaction term in (\ref{afeq3}) is not $-\Lambda $
but $-\Lambda /2$, which is different from (\ref{eqm3}) and (\ref{feq}). The
KGE equations given in (\ref{afeq1})-(\ref{afeq3}) are obtained by just
replacing $\Lambda \rightarrow \Lambda /2$ in all the equations (\ref{eeq1}%
)-(\ref{feq}). In other words, the phase transition makes the scalar field
effectively less self-interactive. This result apparently tells us that
after phase transition the boson star composed of real scalar fields is to
have a smaller mass than in the case before phase transition.

Such a loss in the boson star mass caused by the phase transition, however,
does not appear to be a serious shortcoming for describing the dark matter.
Rather, the fact that the boson star can still exist after a phase transition,
seems to deserve our attention.

\section{Discussions}

In this paper, we have considered self-gravitating solitonic objects made up
of quantum complex/real scalar fields. It has been shown that quantum
complex/real scalar fields may compose the boson stars that are similar to
that formed by classical complex scalar fields. The difference in the
theories considered here appears only in the coefficient of the
self-interaction term in the Einstein equations, i.e. they are $\Lambda /2$
for classical complex fields, $5\Lambda /8$ for quantum complex ones, and $%
3\Lambda /4$ for quantum real ones. Numerically we have verified, Fig.1 and
Fig.2, that the maximum mass of the boson stars increases with the magnitude
of the coefficient of the self-interaction term. This result can be
understood as follows; after eliminating the time-dependent part the KGE
equations for the quantum complex/real scalar fields become effectively for
two ``real'' fields as (\ref{eqmx}) and (\ref{eeq}). Then, the equations
include interactions between the two ``real'' fields as well as
self-interactions of each fields. On the other hand, the KGE equations for
the classical complex fields considered in Ref.\cite{colpi} become
effectively for one ``real'' field after eliminating the time-dependent part
and contain just the self-interaction term of the ``real'' field. Therefore,
the coefficients of the resulting interaction terms in the KGE equations for
the quantum complex/real scalar fields, which are given by (\ref{eqm1}), (%
\ref{eqm2}), and (\ref{eeq1}), (\ref{eq2}), are greater than that in the
case of the classical complex scalar fields. According to the argument that
more interaction energy generates effectively a larger maximum mass of the
boson star, the maximum masses of the boson stars composed of quantum
complex/real scalar fields are greater than that in the case of classical
complex scalar fields. This argument, however, has a limitation due to the
presence of a crossing between the mass-curves in Fig.1 and Fig.2.

In fact, the existence of such solitonic solutions does not guarantee that
the bosonic compact objects can be formed in the universe. (cf. \cite{lopez})
It has been shown
that there is a process that is able to describe the formation of the
bosonic compact objects in the early universe \cite{seidel2}. The process,
which is similar to the way of describing the settling of collisionless star
systems, starts with collapsing due to a gravitational instability analogous
to the Jeans instability. Then, after undergoing the so called gravitational
cooling mechanism ejecting part of the scalar field, bosonic compact objects
could be formed from the primodial bosonic cloud. In the case of complex
scalar fields, this mechanism works to form mini-boson stars. However, in
general, the oscillatons made up of classical real scalar fields can be
formed only in a short dynamical time scale, but are unstable in such a
state. Thus, without introducing additional proviso such as fragmentation of
the primodial bosonic cloud, the classical real scalar fields would be ruled
out as a candidate for the dark matter.

In this paper, we have shown that quantum real scalar fields can form the 
{\it boson star} rather than the {\it oscillaton}. Especially, without the
self-interaction $\Lambda =0$, the mini-boson star composed of the quantum
real scalar fields becomes exactly identical to that made up of the
classical (and quantum) complex scalar fields. Thus, considering the quantum
effect, the real scalar field can be saved, and can be the most promising
candidate for dark matter.

As a corollary, which would be an interesting issue in the study of bosonic
objects, the quantum complex/real scalar fields in excited states may be
considered. The solitonic solutions, if they exist, are unlikely to be
static in both (complex and real) cases. However, it is not obvious if such
quantum fields in excited states can be ruled out completely as candidates
for the dark matter.

Another interesting test for the boson star composed of the quantum real
scalar fields is the second-order phase transition. In this analysis, we
have shown that the phase transition makes the fields effectively more
massive, $m\longrightarrow \sqrt{2}m$, or equivalently less
self-interactive, $\Lambda \rightarrow \Lambda /2$, and more importantly the
boson stars can exist even after the phase transition.

\begin{acknowledgement}

We would like to thank B. Teshima for his prominent contribution to the
numerical job, and S. P. Kim and S. Sengupta for their helpful discussions
and comments. J.H. and C.H.L. also thank to D. Page and V. Frolov for their
warm hospitality during staying at Univ. of Alberta. This work was supported
in part by Korea Science and Engineering Foundation under Grant no.
1999-2-112-003-5 (J.H. and C.H.L.), the BK21 project at Hanyang University
(C.H.L.), and the Natural Science and Engineering Research Council of
Canada(J.H., C.H.L. and F.C.K.).

\end{acknowledgement}

\begin{figure}[htbp]
\begin{center}
    \leavevmode
    \epsfbox{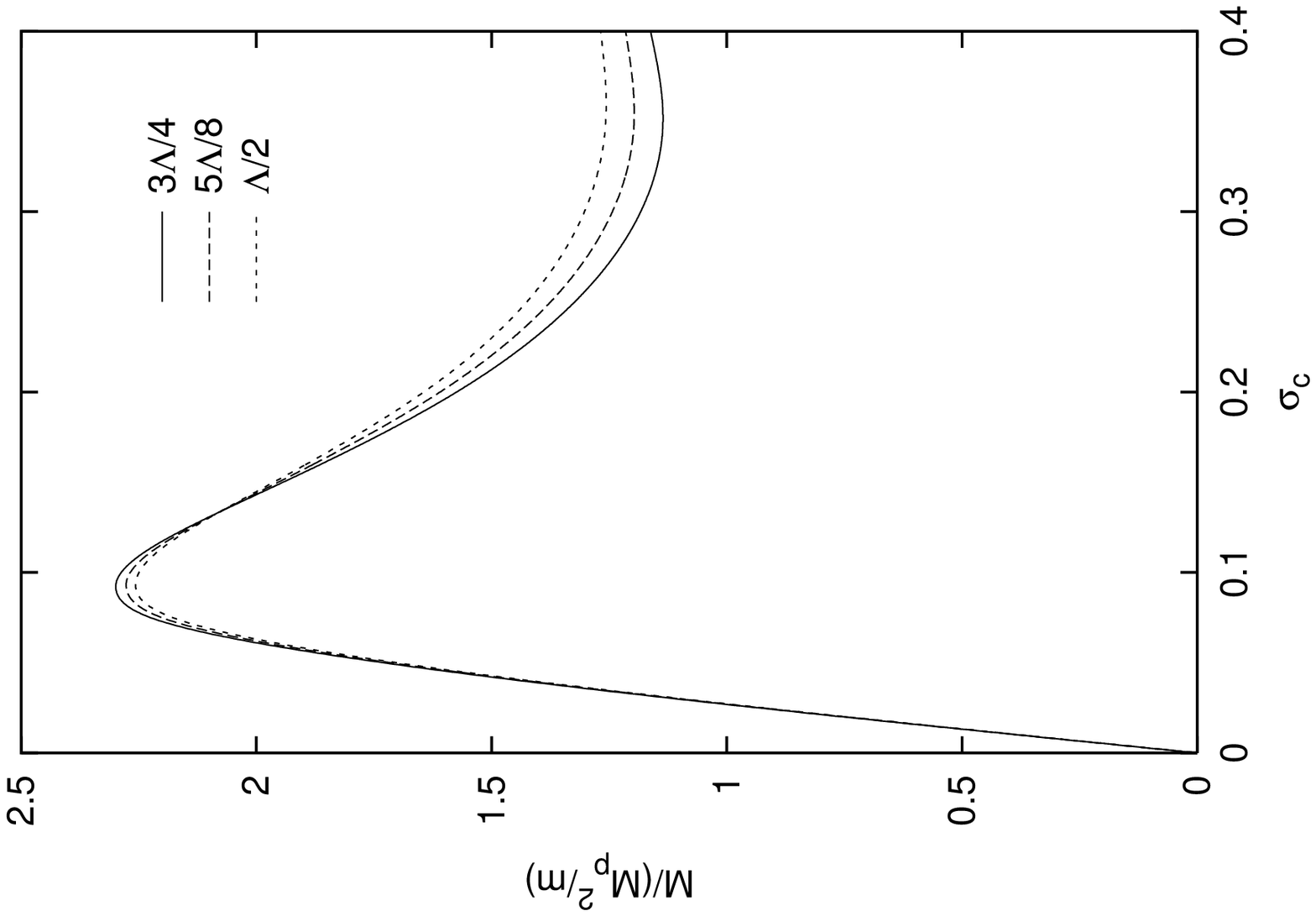}
\end{center}
  \caption{Boson star masses as a function of $\sigma_c$ when $\Lambda=100$. 3$\Lambda$/4, 5$\Lambda$/8, $\Lambda$/2 denote the cases of real and complex quantum scalar fields, and classical complex ones, respectively.}
  \label{fig:fig1}
\end{figure}

\begin{figure}[htbp]
  \begin{center}
    \leavevmode
    \epsfbox{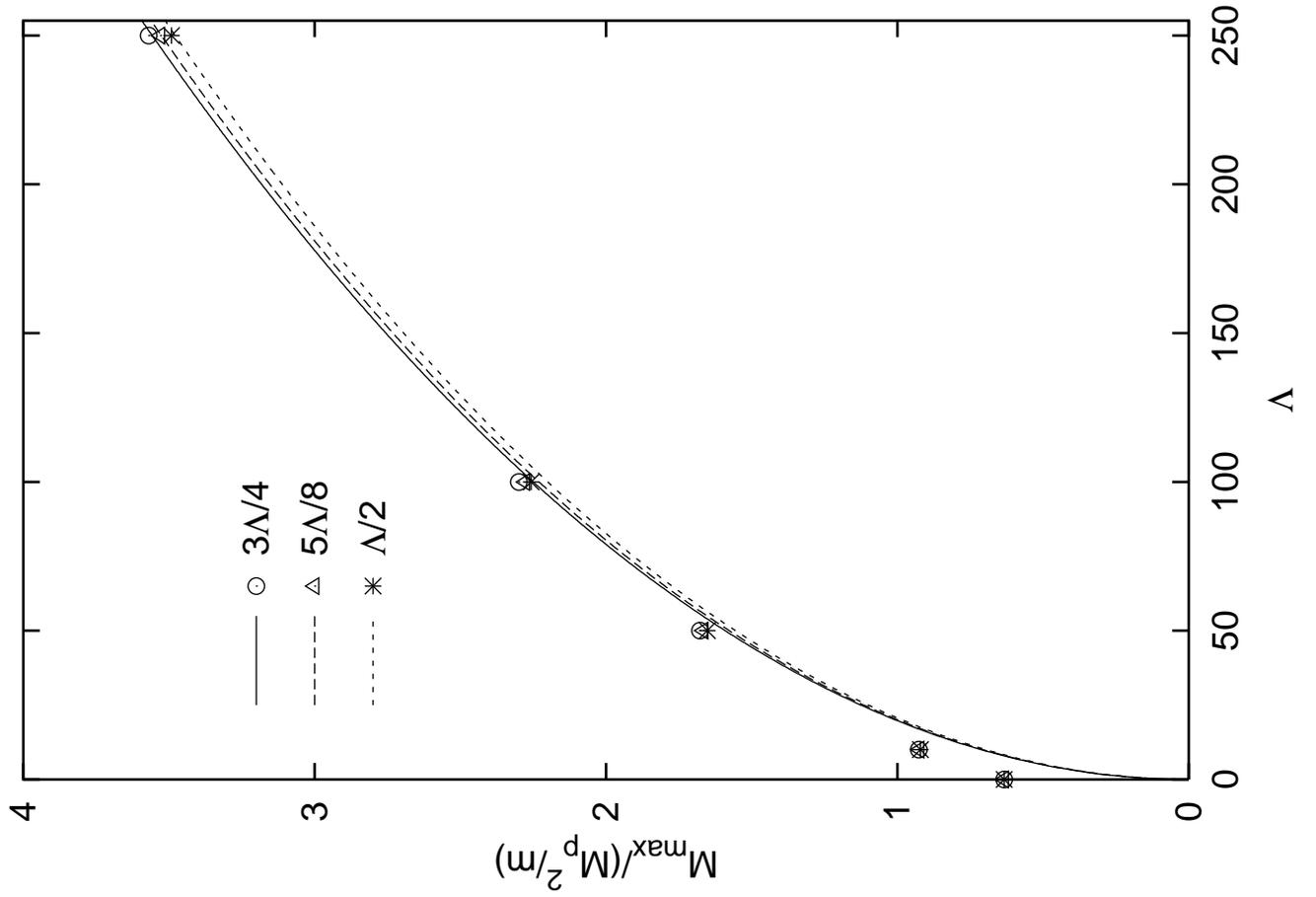}
  \end{center}
  \caption{Maximum masses of boson stars as a function of $\Lambda$. 3$\Lambda$/4, 5$\Lambda$/8, $\Lambda$/2 denote the cases of real and complex quantum scalar fields, and classical complex ones, respectively.}
  \label{fig:fig2}
\end{figure}

\begin{figure}[htbp]
  \begin{center}
    \leavevmode
    \epsfbox{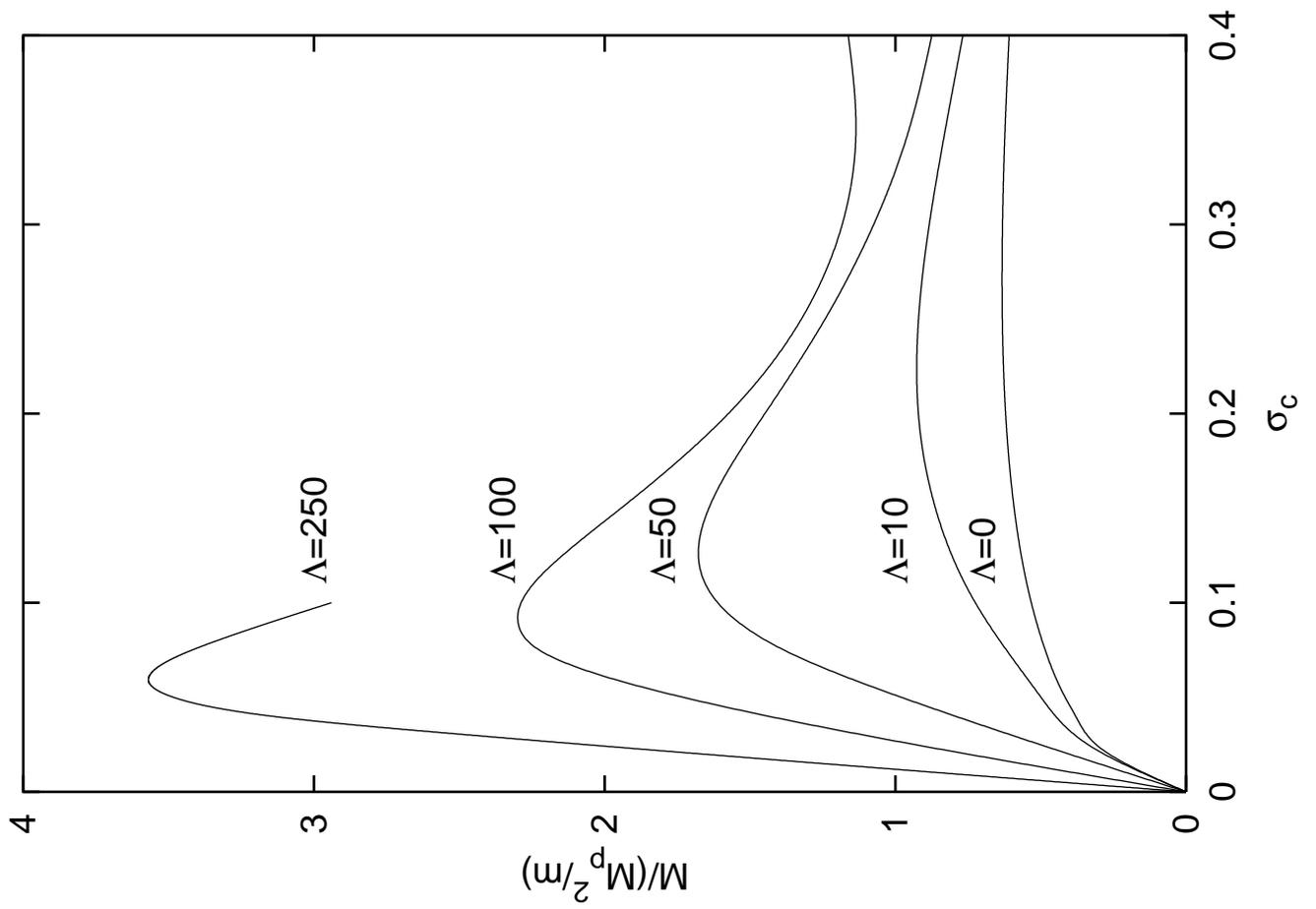}
  \end{center}
  \caption{Mass of boson stars composed of real quantum scalar fields as a function of $\sigma_c$ when $\Lambda=0,10,50,100,250$.}
  \label{fig:fig3}
\end{figure}

\end{document}